\documentclass{article}
\usepackage[T1]{fontenc}
\usepackage[utf8]{inputenc}
\usepackage{ismir} %
\usepackage{amsmath,cite,url}
\usepackage{graphicx}
\usepackage{color}
\usepackage{multirow}
\usepackage{booktabs}
\usepackage{adjustbox}
\usepackage{amssymb}
\usepackage{pifont}  %
\usepackage{enumitem} %
\usepackage{array}      %
\usepackage{listings}
\usepackage{subfigure}
\usepackage{microtype}

\usepackage{xcolor}
\newcommand{\cmark}{\ding{51}} %
\newcommand{\xmark}{\ding{55}} %

\title{CMI-Bench: A Comprehensive Benchmark for Evaluating Music Instruction Following}

\multauthor
  {Yinghao Ma$^{1}$ \hspace{.35cm} Siyou Li$^1$ \hspace{.35cm} Juntao Yu$^1$ \hspace{.35cm}Emmanouil Benetos$^1$ \hspace{.35cm} Akira Maezawa$^2$}
  {{\bf }%
  $^1$ Queen Mary University of London, London, UK\\
  $^2$ Yamaha Corporation, Hamamatsu, Japan\\
  {\tt\small emmanouil.benetos@qmul.ac.uk, akira.maezawa@music.yamaha.com}
  }

\def\authorname{Y. Ma, S. Li, J. Yu, E. Benetos, and A. Maezawa}

\begin{document}

\maketitle

\begin{abstract}
Recent advances in audio-text large language models (LLMs) have opened new possibilities for music understanding and generation. However, existing benchmarks are limited in scope, often relying on simplified tasks or multi-choice evaluations that fail to reflect the complexity of real-world music analysis. We reinterpret a broad range of traditional MIR annotations as instruction-following formats and introduce CMI-Bench, a comprehensive music instruction following benchmark designed to evaluate audio-text LLMs on a diverse set of music information retrieval (MIR) tasks. These include genre classification, emotion regression, emotion tagging, instrument classification, pitch estimation, key detection, lyrics transcription, melody extraction, vocal technique recognition, instrument performance technique detection, music tagging, music captioning, and (down)beat tracking — reflecting core challenges in MIR research. Unlike previous benchmarks, CMI-Bench adopts standardized evaluation metrics consistent with previous state-of-the-art MIR models, ensuring direct comparability with supervised approaches. We provide an evaluation toolkit supporting all open-source audio-textual LLMs, including LTU, Qwen-audio, SALMONN, MusiLingo, etc. Experiment results reveal significant performance gaps between LLMs and supervised models, along with their culture, chronological and gender bias, highlighting the potential and limitations of current models in addressing MIR tasks. CMI-Bench establishes a unified foundation for evaluating music instruction following, driving progress in music-aware LLMs.
\end{abstract}

\vspace{-8pt}
\section{Introduction}\label{sec:introduction}
\vspace{-4pt} 
The emergence of large language models (LLMs) has reshaped the landscape of natural language processing by enabling general-purpose models to solve a wide variety of tasks through instruction following. This paradigm—where models are trained not just on pre-text corpora but instruction-response pairs—has unlocked new possibilities in model generalization, few-shot learning, and cross-domain reasoning. Supervised fine-tuning (SfT), also known as instruction finetuning, and reinforcement learning from human feedback have further strengthened LLMs’ ability to align with human intent%
\cite{zhao2023survey}.

In the context of music, the instruction-following paradigm holds particular promise. Many music-related tasks are naturally multimodal and domain-specific and often lack large-scale annotated data. Instruction-tuned models can generalize to previously unseen problems such as chord generation under rhythmic constraints or personalized music recommendation based on context. Besides, by supporting in-context learning, LLMs offer a flexible path to interact with world music traditions, rare genres, and diverse user preferences—all without explicit retraining\cite{FM4MusicSurvey}.

Recently, a growing number of audio LLMs \cite{liu2024music, deng2024musilingo, chu2023qwen,
tangsalmonn}, extended LLMs with audio encoders and instruction-following capabilities. 
However, these models have so far been evaluated on limited tasks, relying on caption similarity metrics on datasets like\cite{liu2024music, deng2024musilingo}, single-choice protocols \cite{yang2024air, DBLP:conf/ismir/WeckMBQFB24, li2024omnibench}, or multiple-choice question (MCQ) protocols \cite{sakshi2024mmau}
Despite these successes, such evaluations fail to capture the complexity of core music information retrieval (MIR) tasks and offer limited insight to real-world performance.

This work makes three key contributions:
First, we reinterpret a broad range of core MIR annotations as instruction-following tasks as illustrated in Figure \ref{fig:example}, enabling the use of a wide range of MIR datasets, including sequential tasks, not only for evaluation but for training and SFT audio-text LLMs. %
Second, we provide a standardized benchmarking framework that includes implementations of major open-source audio-text LLMs, along with evaluation metrics aligned with prior MIR literature. Unlike earlier MCQ protocols, CMI-bench adopts open-ended, task-specific metrics, allowing more rigorous %
comparisons. 
Last, we present an initial analysis of generalization outside training data, along with cultural and gender bias across models, uncovering potential limitations in their generalization and pointing future directions for culturally inclusive music AI. 
Together, these contributions lay groundwork for systematic progress in music instruction following and its intersection with traditional MIR.
The code\footnote{\url{https://github.com/nicolaus625/CMI-bench/}} and testset audio\footnote{\url{https://huggingface.co/datasets/nicolaus625/CMI-bench}} are available.
\begin{figure}[th]
  \centering
  \includegraphics[alt={ISMIR 2025 template example image},width=\linewidth]{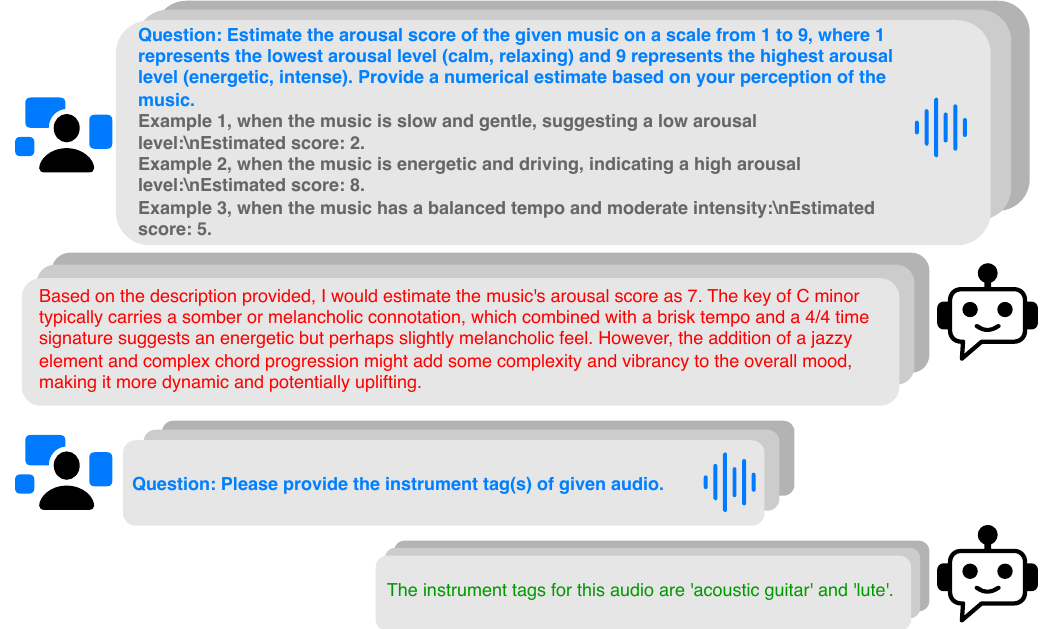}
  \caption{Instruction-Following Format Data in CMI-Bench and example response from Qwen2-audio.}
  \label{fig:example}
\end{figure}

\vspace{-10pt}
This paper is organized as follows. Section 2 reviews related work. %
Section 3 introduces the CMI-Bench design and instruction construction procedure, %
Section 4 describes the experimental setup, including evaluation protocol and open-source models. %
Section 5 presents benchmarking results. %
Finally, section 6 draw the conclusion. %

\section{Related Work}
\subsection{Instruction Following Dataset}
\textbf{Instruction following} refers to the ability of LLMs to perform tasks based on natural language prompts that describe the task itself. This paradigm has become central to recent advances in NLP, where SFT models are trained on a wide range of instruction-response pairs. 
Super-NaturalInstructions\cite{wang2022super} rewrote annotations from over 1,600 diverse NLP tasks into instruction-following formats, showing that models can generalize to unseen tasks given clear instructions. 
Self-Instruct\cite{wang2023self} further advanced this approach by automatically generating diverse instruction–response pairs using the model’s own outputs, while Instructional-GPT\cite{ouyang2022training} aligned models with human intent through SFT and reinforcement learning. %

These techniques have recently been extended to music, enabling instruction-following models to engage with multimodal and domain-specific tasks. MusicQA\cite{liu2024music} and MusicInstruct\cite{deng2024musilingo} repurpose descriptions and tags from MIR dataset to generate Q\&A pairs. 
Such a dataset does not distinguish subtask on instrument, emotion, genre, and caption Q\&A pairs, and the evaluating metrics are BERT-score, overestimate model's music understanding capability without equally compared with traditional MIR algorithms.
Finally, Audio-FLAN\cite{xue2025audio} presents a large-scale instruction-tuning corpus across 80 tasks, unifying understanding and generation in audio, music, and speech. Yet, many tasks are paraphrased to an MCQ format, significantly smaller than the range of pre-defined classes or labels.
Furthermore, these works do not provide a model performance benchmark on such tasks, and the evaluation metrics are not compatible with traditional MIR method.

\subsection{Instruction-Following Benchmarks for Music}
While instruction-following has shown great promise in natural language and vision tasks, its application in the music domain remains underexplored. 
ZIQI-Eval\cite{li2024music} is an instruction following the benchmark on textual symbolic music. 
AIR-Bench\cite{yang2024air} covers a broader range of audio types, including music, but emphasizes low-level tasks such as pitch and instrument recognition and relies primarily on MCQ formats. 
MMAU\cite{sakshi2024mmau} includes music reasoning, yet only covers six MIR datasets, lacks alignment with MIR-specific evaluation metrics, and reports only average scores across tasks. 
MuCho Music\cite{DBLP:conf/ismir/WeckMBQFB24} evaluates music understanding in multimodal models through 1,187 MCQs. %
AudioBench\cite{wang2024audiobench} and MusicBench\cite{melechovsky2024mustango} primarily target audio and text-to-music generation respectively, without addressing MIR tasks. 
MuChin \cite{wang2024muchin}, while valuable for colloquial descriptions, is tailored to Chinese pop song generation. 
Across these efforts, %
most benchmarks omit key MIR tasks popularized by MIREX %
, few support sequential tasks, and evaluation protocols often rely on multiple-choice questions rather than the task-specific metrics used in supervised MIR literature.

\subsection{Audio-Textual Large Language Models}
Current audio-textual LLMs typically consist of an encoder of speech, audio or music, an intermediate architecture %
and an LLM backbone.
LTU\cite{gong2024listen} and LTU-AS\cite{gong2023joint} focus on general audio comprehension and reasoning, combining whisper speech encoder \cite{radford2022robust} w. %
MU-LLaMA\cite{liu2024music}, MusiLingo\cite{deng2024musilingo} and Llark\cite{gardner2024llark} are tailored for music-related tasks, leveraging audio encoders and instructional datasets to support captioning and open-ended music question answering. 
Pengi\cite{deshmukh2023pengi} frames all audio tasks as text generation, unifying audio perception with LLM-based reasoning via a simple prefix-tuning strategy. 
GAMA\cite{ghosh2024gama} and GAMA-IT\cite{ghosh2024gama} integrate multi-layer audio features and instruction tuning (CompA-R) to support complex reasoning over general audio, including music. 
SALMONN-Audio\cite{tangsalmonn} introduces a Q-former window architecture for sequential speech and sound understanding. 
Qwen-Audio\cite{chu2023qwen} and Qwen2-Audio\cite{chu2024qwen2} scale instruction tuning across over 30 audio tasks with hierarchical or natural prompts. %
Audio-Flamingo\cite{kong2024audio} and Audio-Flamingo2\cite{ghosh2025audio} incorporate in-context learning and retrieval-based adaptation for audio-text interaction and dialogue. 
Beyond open-source models, proprietary systems Gemini-2.5 Pro and GPT-4o may represent the state-of-the-art (SOTA). %

\section{CMI-Benchmark}
With CMI-Bench, we aim to address the following limitations in evaluating music understanding capabilities of audio-text LLMs. 
Previous benchmarks often cover only a narrow range of tasks, and no benchmark supports sequential tasks, overlooking many classic challenges which are central to MIR research. 
Moreover, evaluation protocols are typically inconsistent with standard MIR metrics, difficult to compare against traditional supervised models. 
To address these issues, we reformulate annotations from widely-used MIR datasets into instruction-following prompts and process model outputs into formats compatible with standard MIR Python library \texttt{mir\_eval}\cite{raffel2014mir_eval}. 

\subsection{Overview}
\begin{table}[htb]
    \centering
    \renewcommand{\arraystretch}{1.1}
    \resizebox{\columnwidth}{!}{%
        \begin{tabular}{c|ccc}
        \toprule
            Tasks & Dataset & Metrics & \#Test Samples\\ \hline
            Key detection &  GS~\cite{knees2015two} & Gmean score &2406 \\ \hline
            Emotion Regression& EMO~\cite{soleymani20131000} & $R^2$&125\\ \hline
            \multirow{2}{*}{Music tagging} & MagnaTagATune~\cite{law2009evaluation} & ROC-AUC, PR-AUC& 5329\\ 
             & MTG-Top50~\cite{bogdanov2019mtg} & ROC-AUC, PR-AUC &11356\\ \hline
            \multirow{2}{*}{Instrument Classification} &MTG-Instrument~\cite{bogdanov2019mtg}&ROC-AUC, PR-AUC&5115\\ 
             &Nsynth-Instrument~\cite{engel2017neural}&Accuracy&4096\\ \hline
            \multirow{2}{*}{Genre classification} &MTG-Genre~\cite{bogdanov2019mtg}&ROC-AUC, PR-AUC &11479\\
             &GTZAN~\cite{tzanetakis2002musical}&Accuracy&290\\ \hline
             Emotion tagging&MTG-Emotion~\cite{bogdanov2019mtg}&ROC-AUC, PR-AUC & 4231\\ \hline
             Pitch Estimation&Nsynth-Pitch~\cite{engel2017neural}&Accuracy&4096\\ \hline
             Singing Techniques&VocalSet~\cite{wilkins2018vocalset}&Accuracy&1140\\ \hline
             \multirow{2}{*}{Music Captioning} &SDD~\cite{manco2023song}&BL., ME., RO., Bert-Score& 1106\\ 
            &MusicCaps~\cite{agostinelli2023musiclm}&BL., ME., RO., Bert-Score& 2813\\ 
             \hline
             Lyrics Transcription &DSing~\cite{Roa_Dabike-Barker_2019}&WER, CER& 482\\ \hline
             \multirow{2}{*}{Beat tracking} &GTZAN-Rhythm~\cite{tzanetakis2002musical}&F\_measure &290\\ 
             &ballroom~\cite{gouyon2006experimental, krebs2013rhythmic}&F\_measure &685\\  \hline
             \multirow{2}{*}{DownBeat tracking} &GTZAN-Rhythm&F\_measure &290\\ 
             &ballroom&F\_measure & 685\\ \hline
             Melody Extraction &MedleyDB v2~\cite{Bittner2014} &Melody Accuracy& 618\\ \hline
            Performance Technique & GuZheng\_99~\cite{LiCMWYXL23}& frame-level micro/macro-f1&94 \\ 
        \bottomrule 
        \end{tabular}%
    }
    \caption{Overview of tasks, datasets, evaluation metrics, and the number of test samples in the CMI-Bench.}\label{lab:dataset}
\end{table}

The CMI-Benchmark encompasses 14 tasks spanning multi-class, multi-label, regression, captioning, and sequential prediction challenges, evaluated across 20 diverse datasets. 
This benchmark integrates traditional MIR tasks with emerging music-and-language objectives, providing a robust platform to assess computational music intelligence. 
The tasks and datasets used in the benchmark are shown in Table \ref{lab:dataset}. By standardizing splits and metrics, CMI-bench ensures reproducibility and fair comparisons. %

\subsection{Self-Instruction of MIR Annotations}

In this subsection, we introduce the self-instruction framework for CMI-Bench designed to unify diverse MIR tasks under a consistent NLP paradigm, outlining the design of instructions and input tailored to tasks such as key estimation, genre classification, emotion regression, instrument tagging, and temporal sequence annotations.
Our approach leverages structured prompts with multi-class, regression, and sequence-based outputs, enriched with few-shot examples to guide annotation generation. 

For multi-label tasks, we allow flexible outputs without providing pre-defined tags, reflecting real-world complexity.
For clip-level multi-class tasks with a manageable number of categories, such as musical key estimation and genre and vocal techniques classification, instructions explicitly list all possible choices. For instance, key estimation requires selecting one of 24 major and minor keys %
, with few-shot examples like "Bb major" to clarify the format. 
In cases with larger class sets, such as pitch classification on short excerpts across MIDI numbers 9 to 119, we provide a definition of MIDI standard %
alongside examples (e.g., "A4: 69", "Middle C (C4) = 60") to anchor the task.

Regression tasks, such as arousal estimation, adopt a numerical scale (1 to 9) with descriptive anchors—1 for "calm, relaxing" and 9 for "energetic, intense."  To better utilize LLM-s in-context learning capability, %
we include examples for few-shot learning on tie scores to musical characteristics (e.g., "slow and gentle: 2," "energetic and driving: 8"), enabling precise emotional annotation.

Temporal tasks, such as beat tracking and instrument performance technique detection, require structured outputs. Beat tracking outputs timestamps in a comma-separated format (e.g., "0.1s, 1.19s, 2.25s"), while Guzheng (traditional Chinese Kyoto) technique detection uses a Python-style list of tuples (e.g., "[('70.8086', '71.4817', 'Tremolo')]"), covering techniques like Vibrato and Glissando. Default outputs "[('0.0', '10.0', 'No Tech')]" handle cases with no detections. 
Melody extraction follows similar principles, balancing specificity and clarity. 
We forbid tuples to have time overlapping on melody and (down)beat tracking, but allow for playing technique detections.

Inputs are uniformly represented as audio placeholders (e.g., "<|SOA|><AUDIO><|EOA|>"), paired with metadata such as audio paths and time segments. This ensures compatibility with NLP models while preserving MIR task diversity, offering a scalable framework for future efforts.

\section{Experiments}
\subsection{Evaluation Protocol}
To enable rigorous and fair comparison with traditional MIR systems, we design an evaluation pipeline that closely follows the original task definitions and metrics. 
All model outputs are automatically post-processed to conform to each task’s expected format, ensuring compatibility with MIR evaluation tools such as \texttt{mir\_eval}. Below, we detail the evaluation strategies used for each task category.
\subsubsection{Classification Tasks}
\textbf{Multi-Class Classification}. Tasks include short-clip monophonic pitch estimation, instrument classification, singing technique classification, and genre classification. We evaluate using strict string matching: a model’s response is considered correct if it contains only the correct label (case-, space-, and punctuation-insensitive) and no others. For pitch classification, we additionally require the model to follow the instruction format and return MIDI numbers. Accuracy is used as the metric.

\textbf{Multi-Label Classification} Tasks include music tagging, genre labelling, emotion tagging, and instrument recognition. As model responses may include synonyms or free-form text, we embed both the predicted and ground truth tag sets using the BGE encoder \cite{xiao2024c}, a model optimized for retrieval and multi-label matching. Cosine similarity scores are then used to compute ROC and PR, providing a soft but semantically aligned evaluation quality.
\subsubsection{Clip-level MIR Tasks}
\textbf{Key Detection}. We adopt standard weighted score metric from \texttt{mir\_eval.key}, which accounts for musically reasonable errors, such as relative minor or parallel key. %

\textbf{Regression}. Model outputs are constrained to integers in the range [1, 9]; if a float is returned, we take the floor. Outputs are then z-score normalized to zero mean and unit variance. If a model fails to return a value, we assign the model’s mean value. The coefficient of determination ($R^2$) is computed between predictions and annotations on arousal and valance.

\textbf{Music Captioning.}
We assess caption quality using four standard NLP metrics: BLEU~\cite{Papineni02bleu:a, lin-och-2004-orange}, METEOR~\cite{banerjee2005meteor}, ROUGE~\cite{lin-2004-rouge}, and Bert-Score~\cite{bert-score}. %

\subsubsection{Sequential MIR Tasks}
\textbf{Lyrics Transcription}.
We extract lyrics from model outputs by removing typical prefixes (e.g., “lyrics is as follows:”). Word Error Rate (WER) and Character Error Rate (CER) are computed against ground-truth lyrics.

\textbf{(Down)Beat Tracking.}
Model are expected to return a list of time points for (down)beat events. We filter non-numeric outputs, sort the list by time, and apply F-measure metric from \texttt{mir\_eval.beat}, with a 20ms tolerance. %

\textbf{Melody Extraction} is treated as a sequential regression task on the fundamental frequency of notes calculated by \texttt{mir\_eval.melody.evaluate} with 50 music cents tolerance.
Models are instructed to return a list of (time, pitch) tuples. We discard invalid tuples (e.g., missing pitches, or improperly formatted entries, etc.). If multiple pitches are predicted for the same timestamp, we use only the first. Evaluation is based on frame-level accuracy.

\textbf{Instrument Playing Technique Detection}.
For the GuZheng\_99 dataset, we evaluate frame-level predictions using macro- and micro-F1 scores, allowing for overlapping techniques. 
Invalid predictions (e.g., incorrect tuple formats) are filtered out. 
Empty responses are interpreted as a "no technique" prediction covering the full time range.

\subsection{Models}

\begin{table}[htb]
\centering
\resizebox{\columnwidth}{!}{%
\begin{tabular}{lcccc}
\toprule
\textbf{Model} & \textbf{\#Params} & \textbf{Sound} & \textbf{Music} & \textbf{Speech}\\& & \textbf{Encoder} & \textbf{Architecture} & \textbf{Decoder} \\
\midrule
Pengi~\cite{deshmukh2023pengi} & 323M & \cmark & \cmark & \\ %
Audio-Flamingo~\cite{kong2024audio} & 2.2B & \cmark & \cmark & \\
LTU~\cite{gong2024listen} & 7B & \cmark & \cmark & \\
LTU-AS~\cite{gong2023joint} & 7B & \cmark & \cmark & \cmark  \\
MusiLingo-long~\cite{deng2024musilingo} & 7B &  & \cmark &  \\
MuLLaMa~\cite{liu2024music} & 7B &  & \cmark &    \\
GAMA~\cite{ghosh2024gama} & 7B & \cmark & \cmark &  \\
GAMA-IT~\cite{ghosh2024gama} & 7B & \cmark & \cmark &  \\
Qwen-Audio-Chat~\cite{chu2023qwen} & 8.4B & \cmark &  & \\
Qwen2-Audio-Instruct~\cite{chu2024qwen2} & 8.4B & \cmark & \cmark & \cmark \\
SALAMONN-Audio~\cite{tangsalmonn} & 13B & \cmark & \cmark & \cmark \\
\bottomrule
\end{tabular}%
}
\caption{Comparison of audio-textual LLMs by training domains.%
\cmark denotes coverage or presence; \xmark absence.}
\label{tab:llm_audio_models}
\end{table}

To provide a broad and representative evaluation, we implement and benchmark 11 audio-text LLMs with publicly available weights demonstrated in table \ref{tab:llm_audio_models}. 
Our selection covers a wide spectrum of model designs and training corpus, enabling a comprehensive comparison of instruction-following capabilities across various music-specific tasks.

\section{Results and Discussion}
\subsection{Benchmarking Results}

\begin{table*}[tbh]
\vspace{-25pt}
\centering
\begin{adjustbox}{max width=\textwidth}
\begin{tabular}{@{\hskip 4pt}l|l|ccccccccccc|c@{\hskip 4pt}}
\toprule[1.5pt]
\textbf{}     & \textbf{} & \textbf{Qw2.} & \textbf{Qw.} & \textbf{Salm.} & \textbf{MusiL.} & \textbf{LTU} & \textbf{LTU-AS} & \textbf{MU-L.} & \textbf{auFla.} & \textbf{Gama} & \textbf{GamaI} & \textbf{Pengi} &\textbf{SOTA} \\
\midrule[1pt]
\textbf{GS-K} & GES $\uparrow$      & 8.28 & 6.51 & 7.70 &\textbf{9.50} & 7.61& 1.42 & 7.56 & {8.21} & 7.69 & 7.70& 0.00 & 74.3\cite{korzeniowski2017end} \\
\midrule
\multirow{2}{*}{\textbf{EMO}} & aR2 $\uparrow$ & -0.75 &  -0.44 & -0.51 & -0.68 &  -1.14 & -1.27 &  -0.03 & -0.85 & -1.08 & -0.29 & \textbf{0.00} &0.62\cite{castellon2021codified} \\
                               & vR2 $\uparrow$ & -0.84 & -0.78 & 0.0 &  -0.60 & -1.13 & -0.78 & -0.12 & -0.60 & -1.30 & \textbf{-1.19} & \textbf{0.00} &0.76\cite{yuan2023marble} \\
\midrule
\multirow{2}{*}{\textbf{MTT}} & ROC $\uparrow$ & 66.78 & 66.00 & 59.07 & 63.39 & 65.75 & 65.83 & 68.32 & 68.68 & \textbf{81.21} & 78.32 & 66.75 &92.0\cite{huang2022mulan}\\
                               &\ PR $\uparrow$ & 19.15 & 16.99 & 15.08 & 12.25 & 17.78 & 15.72 & 18.65 & 20.16 & \textbf{34.26} & 27.53 & 17.82 &41.4\cite{castellon2021codified} \\
\midrule
\multirow{2}{*}{\textbf{M-G}} & ROC $\uparrow$ & 64.44 & \textbf{66.39} & 57.71 & 57.48 & 52.22 & 57.14 & 57.36 & 62.83 & 52.50 & 62.49 & 58.23 &88.0\cite{yuan2023marble} \\
                               &\ PR $\uparrow$ & \textbf{9.23}  & 8.07  & 5.62  & 4.99  & 3.62  & 4.98  & 4.97  & 6.85  & 3.90  & 6.01  & 5.47 &20.5 \cite{yuan2023marble}  \\
\midrule
\multirow{2}{*}{\textbf{M-E}} & ROC $\uparrow$ & \textbf{60.89} & 59.06 & 50.69 & 53.07 & 51.41 & 52.02 & 54.40 & 55.80 & 51.97 & 58.84 & 53.88 &78.6\cite{mccallum2022supervised} \\
                               &\ PR $\uparrow$ & \textbf{7.85}  & 6.09  & 3.65  & 3.95  & 3.98  & 3.72  & 4.35  & 4.60  & 4.07  & 5.27  & 3.93 &16.1\cite{mccallum2022supervised} \\
\midrule
\multirow{2}{*}{\textbf{M-I}} & ROC $\uparrow$ & \textbf{58.90} & 56.95 & {48.78} & 55.63 & 55.34 & 53.02 & 50.81 & 56.99 & 51.15 & 55.16 & 56.09 &78.8\cite{alonso2022music} \\
                               &\ PR $\uparrow$ & \textbf{12.41} & 11.35 & 7.44  & 9.24  & 10.98 & {8.90}  & {8.24}  & 10.71 & 9.01  & 10.69 & 9.36 & 22.0\cite{yuan2023marble}  \\
\midrule
\multirow{2}{*}{\textbf{M-50}} & ROC $\uparrow$ & 64.64 & 63.00 & 53.46 & \textbf{57.58} & \textbf{53.86} & 54.11 & 54.88 & 60.96 & 52.01 & 60.68 & 57.22 &84.3\cite{mccallum2022supervised} \\
                                &\ PR $\uparrow$ & 16.54 & 14.45 & 9.49  & 9.68  & 8.30  & 8.67  & 9.11  & 12.16 & 8.10  & 11.72 & 10.19 &32.1\cite{mccallum2022supervised} \\
\midrule
\textbf{GTZ.} & Acc. $\uparrow$ & \textbf{72.07}  & {71.38} & 32.76 & 7.24 & 2.76 & 16.90 & 8.97 & 50.34 & 21.38 & 42.41 & 6.21 &83.9\cite{niizumi2023masked}\\ \hline

\textbf{VS-T} & Acc. $\uparrow$ & {14.91} & 15.18 & \textbf{15.61}  & 1.23 & 7.11  & 0.53& 4.56 & 11.32  &{7.72}& 7.89  & 0.00 &76.9\cite{li2023mert}  \\ \hline

\textbf{NI}   & Acc. $\uparrow$ & 37.62 & 4.13  & 0.15  & 0.00  & 0.49 & 6.88 & 0.00   & 15.80 & \textbf{58.37} & 39.36 & 42.26 & 78.2\cite{wang2022towards}] \\ \hline
\textbf{NP}   & Acc. $\uparrow$ & 1.51  & 0.37  & 0.00   & 0.00  & 0.73 & 0.05 & 0.00   & 0.73  & 0.20  & 0.00   & \textbf{5.74} & 89.2\cite{mccallum2022supervised} \\
\midrule
\multirow{4}{*}{\textbf{SDD}} & BL. $\uparrow$     & \textbf{23.40}         & 11.95        & 16.41          & 8.14            & 11.54        & 9.72            & 15.55        & 15.14           & 15.96        & 20.93          & 15.47     & -     \\
                    & ME. $\uparrow$              & \textbf{23.21}         & 9.35         & 18.45          & 14.32           & 8.51         & 7.49            & 13.89        & 11.81           & 13.81        & 16.41          & 9.98     &16.7\cite{doh2023lp}      \\
                    & RO. $\uparrow$             & \textbf{28.47}         & 12.35        & 28.12          & 30.15           & 9.33         & 9.42            & 15.28        & 12.92           & 18.35        & 20.07          & 11.45      &111.9\cite{doh2023lp}    \\
                    & BS. $\uparrow$             & \textbf{87.44}        & 84.79        & 86.68          & 85.28           & 84.44        & 83.62           & 86.38        & 85.75           & 85.89        & 86.21          & 82.90     &86.0\cite{doh2023lp}     \\ 
\midrule
\multirow{4}{*}{\textbf{MC}}  & BL. $\uparrow$     & 14.76         & 2.98         & 1.23           & \textbf{21.50}           & 5.24         & 4.22            & 3.48         & 2.25            & 7.57         & 14.53          & 16.52     &21.7\cite{deng2024musilingo}     \\
                    & ME. $\uparrow$              & 12.47         & 5.55         & 4.60           & \textbf{22.49}           & 8.55         & 7.01            & 8.01         & 5.97            & 10.07        & 10.98          & 14.77       &22.4\cite{doh2023lp}   \\
                    & RO. $\uparrow$              & 12.35         & 6.68         & 6.26           & \textbf{30.29}           & 9.39         & 7.51            & 8.58         & 6.94            & 11.38        & 12.46          & 12.64   &30.8\cite{deng2024musilingo}       \\
                    & BS. $\uparrow$               & 84.38         & 82.37        & 82.98          & \textbf{85.75}           & 83.84        & 83.59           & 83.00        & 83.43           & 84.30        & 84.57          & 83.22    &87.8\cite{doh2023lp}      \\ 
\midrule
\multirow{2}{*}{\textbf{DS}}         & WE. $\downarrow$                 & 793.0        & \textbf{115.7}       & 816.1         & 2019         & 235.5       & 191.7          & 191.9       & 275.7          & 225.4       & 152.6         & 343.2     &12.99\cite{ou2022transfer}    \\
                    & CE. $\downarrow$                 & 818.6        & \textbf{96.2 }       & 760.00        & 2311         & 210.8       & 185.5          & 168.3       & 262.6          & 201.3       & 165.2         & 368.0     & -        \\ 
\midrule
\textbf{G-B}        & FM. $\uparrow$               & 7.50          & \textbf{23.69}        & 11.49          & 0.04            & 0.10         & 0.00             & 0.71         & 3.96            & 0.00        & 1.49           & 0.00     &88.3\cite{li2023mert}     \\ 
\midrule
\textbf{G-D}        & FM. $\uparrow$              & 5.97          & \textbf{10.21}        & 8.62           & 0.18            & 0.86         & 0.00             & 0.17         & 3.06            & 0.05         & 0.54           & 0.00   &54.1\cite{DBLP:conf/ismir/HeydariCD21}   \\ 
\midrule
\textbf{BR-B}       & FM. $\uparrow$                & 7.12          & \textbf{21.96}        & 14.97          & 0.01            & 0.15         & 0.00             & 0.22         & 4.69            & 0.02         & 1.02           & 0.00     &96.8\cite{DBLP:conf/ismir/ZhaoXW22}     \\ 
\midrule
\textbf{BR-D}       & FM. $\uparrow$               & 5.69          & \textbf{10.68}        & 9.40           & 0.06            & 2.29         & 0.00             & 0.15         & 3.47            & 0.14         & 0.68           & 0.00      &94.1\cite{DBLP:conf/ismir/ZhaoXW22}    \\ 
\midrule
\textbf{MDB}                   &Acc. $\uparrow$ & \textbf{5.06 }         & 0.08         & 0.00            & 0.00             & 0.00            & 0.00             & 0.00            & 0.01            & 0.66          & 0.00            & 0.00          &72.3\cite{yu2024hkdsme}  \\ \hline
\multirow{2}{*}{\textbf{GZ}}                    & {maF1} $\uparrow$ & \textbf{3.18}         & 1.66         & 0.03           & 0.00             & 0.04           & 0.00             & 0.00            & 0.00             & 0.00           & 0.00            & 0.00        &90.0\cite{li2024mertech}    \\
                               & {miF1} $\uparrow$ & \textbf{0.89}          & 0.44         & 0.01           & 0.00             & 0.01           & 0.00             & 0.00            & 0.00             & 0.00           & 0.00            & 0.00       &80.4\cite{li2024mertech}     \\
\toprule[1.5pt]
\end{tabular}
\end{adjustbox}
\vspace{-12pt}
\caption{Performance of 11 open-source audio-text LLMs on CMI-Bench. Models: Qwen2-Audio (Qw2.), Qwen-Audio (Qw.), SALMONN-Audio (Salm.), MusiLingo (MusiL.), 
LTU, LTU-AS, MU-LLaMA (MU-L.), Audio-Flamingo (auFla.), GAMA, GAMA-IT (GamaI), Pengi. Tasks include key detection (GS-K), emotion regression (EMO), tagging (MTT, M-50), genre (M-G, GTZ.), emotion/instrument tagging (M-E, M-I), captioning (SDD, MC), lyrics transcription (DS), beat/downbeat tracking (G-B/G-D, BR-B/BR-D), melody (MDB), and Guzheng techniques (GZ). Metrics: GES, R², ROC-AUC, PR-AUC, Accuracy, BLEU (BL.), METEOR (ME.), ROUGE (RO.), BERTScore (BS.), WER/ CER, FM(F-Measure), Macro-F1 (maF1), Micro-F1 (miF1). Best scores are in bold.
}\label{tab:optimized-table}
\vspace{-13pt}
\end{table*}

Experiment results reveal several important observations about the current state of audio-text LLMs on MIR tasks.

\subsubsection{LLMs Underperform Traditional MIR Baselines.}
Despite LLMs have achieved excellent results on music captions and multi-choices QA,
\cite{deng2024musilingo, agostinelli2023musiclm, li2024omnibench, DBLP:conf/ismir/WeckMBQFB24,  sakshi2024mmau},
all models in our study fall significantly short of the performance achieved by task-specific supervised systems when evaluated using standard MIR metrics besides music captioning.
This is consistent across classification, regression, and sequential tasks. These findings suggest that instruction-following LLMs still lack the specialized precision and inductive bias of MIR models trained explicitly for each task.

\subsubsection{Best Performance May Skew toward Training Set}
Interestingly, the peak performance on each task is typically achieved by models whose datasets overlap significantly with their training corpus, revealing limited generalization. %
Qwen2-Audio performs best on MTG-Jamendo-related tasks such as MTG-top50, MTG-Emotion, and SDD captioning, while common on other tagging and caption datasets. This aligns with its use of MTG-Jamendo and FMA during model development via AIR-Bench, suggesting unsatisfying generalization capability. %
Besides, MusiLingo performs best on MusicCaps, the same dataset it was trained on for captioning and Q\&A.
Lastly, GAMA shows the best on MTT and NSynth-instrument and comparative on MusicCaps, while common on other datasets on same tasks, reflecting bias in its SFT corpus. %
These demonstrate that supervised instruction-tuned models can capture task-specific patterns well when training data is directly aligned, but their generalization to unseen or structurally different tasks remains limited. %

\subsubsection{All Models Perform Poorly on DSing Transcription}
Despite the absence of instrumental accompaniment and use of English lyrics, none of the models reach usable performance levels on DSing for lyrics transcription, though it is relatively clean. 
This result is particularly striking for models like LTU and SALMONN, which include Whisper as their audio encoder and could theoretically benefit from ASR capabilities. LyricWhiz \cite{zhuo2023lyricwhiz} utilizes GPT-4 to post-process whisper ASR output on DSing dataset, providing results similar to SOTA without training.

\subsubsection{Prompting Format May Impacts Performance.}
Prompting without task-specific tokens used during training significantly degrades performance.
Qwen-Audio performs far worse on Nsynth-Pitch than reported in its original paper. This is likely due to the absence of structured task tokens (e.g., ``<|pitch|><|midi\_pitch|>piano'') in our prompt. Instead, CMI-bench relies on general natural language instructions. 
This highlights a critical gap in current audio LLMs: without clearly defined prompting schemas, their ability to interpret instructions can be fragile and fail to generalize. While different prompts for MusiLingo do not provide a significant difference on MusicCaps.

\subsubsection{Sequential Tasks Remain Challenging for All.}
Tasks involving structured sequence-based outputs—such as melody extraction, instrument performance technique detection, and (down)beat tracking—are poorly handled by all evaluated models. Even Qwen-Audio, which shows relatively strong performance in genre and beat tracking, falls far short of MIR baselines, sometimes copying the input examples. We hypothesize two key reasons:
For one thing, the diversity and ambiguity in how sequence tasks are phrased (e.g., timestamps, tuple formats) reduces consistency in model outputs.
For another, many models have only limited exposure to audio tasks with dense temporal supervision. If pretraining data includes timestamped output and matched decoding formats, performance may improve.

\subsubsection{Emotion Regression Fails for All Models.}
Despite clear instructions, carefully designed scales (1–9), and contextual music descriptions, and few-shot examples, all models fail to provide usable predictions for arousal and valence. In fact, model outputs often cluster around meaningless values, sometimes performing worse than simply predicting the mean. Our post-processing rules convert empty or invalid outputs to dataset means, which often lead to better R² scores than the models themselves—highlighting the severe limitations in mapping continuous perceptual attributes from music using current audio-text LLMs.

These findings emphasize the gap between current SFT multimodal LLMs and traditional task-specific MIR systems. 
While open-source audio LLMs show promise in isolated tasks with aligned training data, substantial challenges remain in terms of generalization, structured output generation, and adaptation to real-world settings.

\subsection{Culture and Gender Bias}

We further analyze the performance of two top-performing models—Qwen2-Audio and Audio-Flamingo—on fine-grained instrument, genre, and music tag categories. While both models show competitive results overall, our breakdown highlights notable performance disparities across instrument types, cultural genres, and voice-related tags.

\begin{figure}[htb]
  \centering
  \includegraphics[alt={ISMIR 2025 template example image},width=\linewidth]{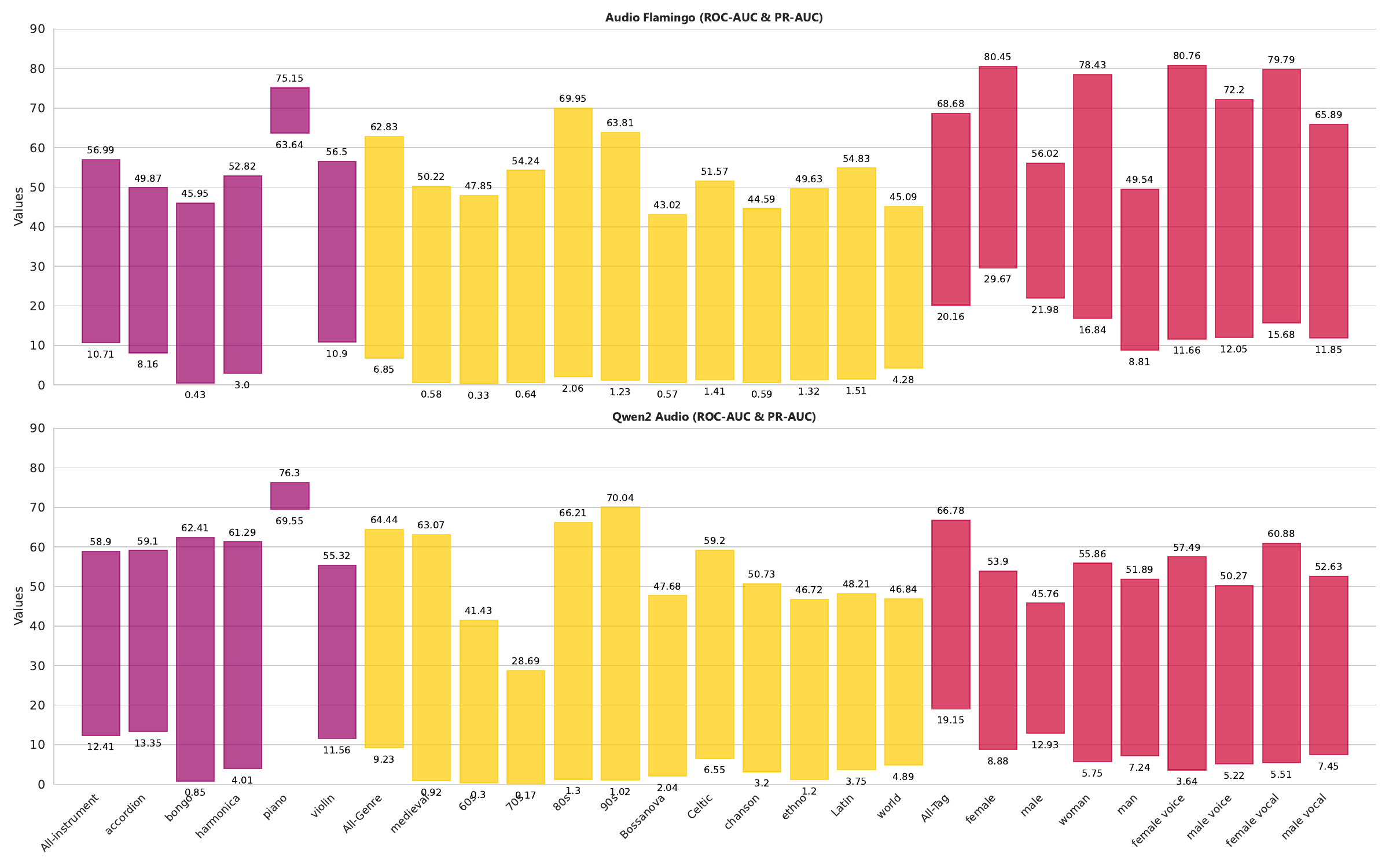}
  \caption{Fine-grained evaluation of Qwen2-Audio and Audio-Flamingo on instrument (purple), genre (yellow), and vocal (red) tag classification. The upper extremity represents the ROC-AUC value, and the lower is PR-AUC.}
  \label{fig:example}
\end{figure}

\subsubsection{Instrument Bias on MTG-Instrument}
Both models achieve high scores on piano, reflecting the strong representation of piano in most training datasets. Western instruments such as violin and accordion perform close to the average, suggesting moderate robustness across common musical timbres. However, performance drops significantly on bongo and harmonica — commonly associated with world music. %
These results point to a persistent bias toward Western instruments and limited generalization to underrepresented timbres in current pre-training corpora.

\subsubsection{Cultural Genre Imbalance on MTG-Genre}
Genre classification results similarly reveal systematic disparities. Both models show relatively strong performance on mainstream Western pop genres (e.g., 80s, 90s), while genres associated with world music (e.g., Bossanova, Celtic, Chanson, Ethno, Latin) and music traditions (e.g., Medieval) consistently fall below average. 
For example, Audio-Flamingo's performance on Bossanova and Chanson drops severely. Qwen2-Audio performs slightly better on some long-tail genres, but still shows considerable degradation. These highlight a lack of cultural and historical diversity in the data used for instruction tuning and model pretraining.

\subsubsection{Voice Tag Differences on MTT}
A detailed comparison on vocal tags reveals an interesting divergence. Audio-Flamingo is consistently better at identifying *female* voices than male voices, indicating a possible gender-related acoustic or annotation bias. In contrast, Qwen2-Audio achieves higher ROC-AUC for *female* tags but lower PR-AUC, suggesting that while the model ranks positive examples correctly, its absolute predictions remain sparse or overconfident. This mismatch implies that Qwen2-Audio is sensitive to class ranking but may lack calibration in estimating tag presence probabilities, an issue worth investigating for fairness and reliability in music model deployment.

\subsection{Ablation Study on Different Prompts and Trials}
\vspace{-2mm}
\begin{figure}[ht]
    \centering
    \includegraphics[width=.7\linewidth]{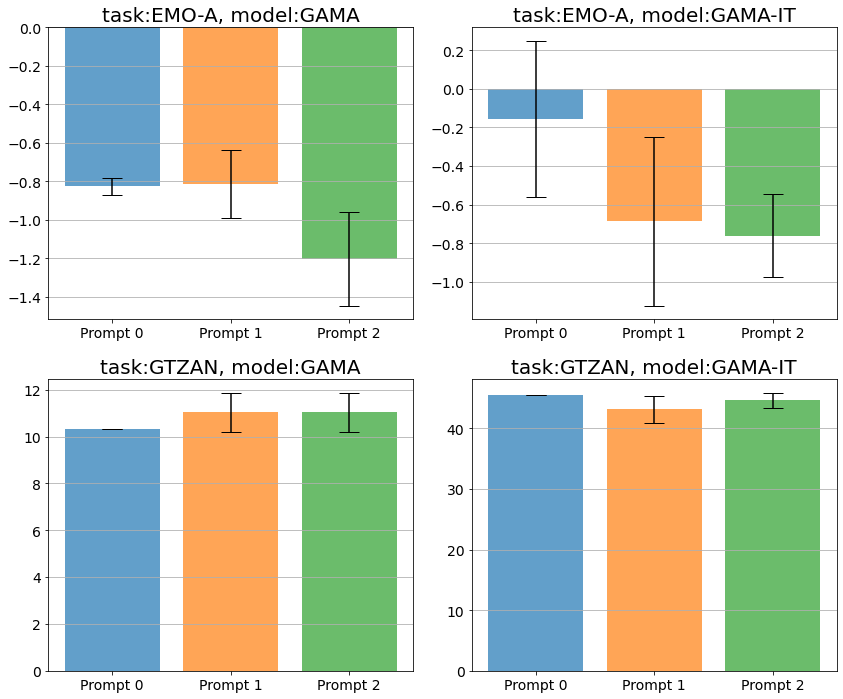}
    \caption{Ablation Study on Prompt Sensitivity for Genre Classification and Arousal Regression}
    \label{fig:enter-label}
\end{figure}
\vspace{-2mm}

We conduct an ablation study on prompt design using GAMA and GAMA-IT models across two representative tasks: GTZAN genre classification and EMO arousal regression. 
Variant Prompts 1 and 2 are evaluated over three runs, and the bars report mean performance with standard variant as error bars.
GTZAN results (bottom row) are relatively stable across prompts and have small variance for each prompt in multi-trials, indicating that most genre-related instructions are consistently followed. The low variance suggests robustness to prompt changes.
In contrast, EMO-A results (top row) show relative sensitivity to prompt variation, particularly under the GAMA-IT model. This instability stems from a higher rate of invalid or non-responsible generations, which are scored as mean values during evaluation. Consequently, differences in prompt phrasing might lead to large deviations, especially when valid predictions diverge significantly from the mean score.

\section{Conclusion}
We introduce CMI-Bench, a comprehensive benchmark for evaluating audio-text LLMs across diverse MIR tasks. Our results highlight a significant performance gap between LLMs and supervised MIR systems, with best models like Qwen2-Audio and GAMA also struggling with generalization. Sequence-based tasks, such as melody extraction and beat tracking, pose particular challenges, likely due to limited timestamped pretraining and prompt sensitivity. Fine-grained analysis also reveals cultural and gender biases tied to training data imbalances. By offering a standardized evaluation framework and toolkit, CMI-Bench bridges NLP and MIR research, providing a foundation for future advancements. Progress will hinge on improved pretraining, sequential output handling, and bias mitigation, and we hope this work spurs collaboration toward more capable music-aware LLMs.

\appendix
\section{Acknowledgments}
Yinghao Ma is a research student at the UKRI Centre for Doctoral Training in Artificial Intelligence and Music, supported by UK Research and Innovation [grant number EP/S022694/1]. 

Siyou Li is a research student at the Computational Linguistics Lab at Queen Mary University of London, funded by the QMUL-CSC PhD scholarships.

Yinghao Ma would also like to express heartfelt gratitude to the Student Philharmonic Chinese Orchestra at the Chinese Music Institute, Peking University (abbreviated as CMI, unrelated to the paper title). We warmly congratulate the orchestra on its 20th anniversary.
\section{Ethics Statement}
CMI-Bench repurposes existing publicly available datasets in the MIR domain by reformatting their annotations into instruction-following formats. No new human annotations were collected, and no human participants were involved in the creation of this benchmark. 
All data used in the project are licensed under terms that permit non-commercial research use. In compliance with these terms, we license CMI-Bench under a Creative Commons Attribution-NonCommercial-ShareAlike (CC BY-NC-SA) license. To promote long-term accessibility, we host the audio test set on Hugging Face with clear usage restrictions for non-commercial purposes.

The dataset primarily consists of Western, English-language popular music, with limited inclusion of instrumental tracks and non-English songs. Transcription tasks are restricted to English lyrics, and world music instruments besides Guzheng are underrepresented. We acknowledge this cultural and linguistic skew and encourage future extensions to improve global diversity and representation.

This work involves no safety, security, or environmental risks. The benchmark does not require high-compute model training or deployment of potentially harmful generative models. We release CMI-Bench and its evaluation toolkit to foster responsible and reproducible research in audio-language modeling.

\section{Appendix}
Due to the limitation of the ISMIR proceeding, please refer to our arxiv version for more information on instruction examples, error case analysis and more discussion.

\bibliography{ISMIRtemplate}

\newpage
\section{Supplementary material \& Appendix}
\subsection{Examples of Instructions}
We change the few-shot examples on the output formats to text descriptions. The updates instructions are as follows.
\begin{lstlisting}
{"instruction": "Please provide the description of given song.", 
"input": "<|SOA|><AUDIO><|EOA|>",
"output": "Uplifting English rock and roll song with a positive feeling.", 
"uuid": "", "split": ["test"], 
"task_type": \begin{lstlisting} {"major": ["captioning"], 
"minor": ["music_captioningn"]}, 
"domain": "music", 
"audio_path": ["data/SDD/audio/audio/71/1009671.2min.mp3"], 
"audio_start": 0.0,
"audio_end": 30.0, 
"source": "MTG-SDD", 
"other": \begin{lstlisting} {"tag": "null"}}
\end{lstlisting}

\begin{lstlisting} 
{"instruction": "You are given an audio clip of a music piece.
Estimate the musical key of the audio by selecting exactly one key from the following list:

C major, Db major, D major, Eb major, E major, F major, Gb major, G major, Ab major, A major, Bb major, B major,
C minor, Db minor, D minor, Eb minor, E minor, F minor, Gb minor, G minor, Ab minor, A minor, Bb minor, B minor.

Return only the name of the predicted key. No explanation.
", "input": "<|SOA|><AUDIO><|EOA|>", 
"output": "E major",
"uuid": "", 
"split": ["test"], 
"task_type": {"major": ["multi-class"], "minor": ["key_detection"]}, 
"domain": "music", "audio_path": ["data/GS-key/giantsteps_clips/wav/4693198-2.wav"], "audio_start": 0.0, 
"audio_end": 30.0, 
"source": "GiantSteps", 
"other": {"tag": "null"}}
\end{lstlisting}
\begin{lstlisting} 
{"instruction": "Arousal refers to the level of physiological or psychological activation a piece of music induces in a listener. It reflects how calming or energizing the music feels.
You are given a piece of music. Estimate its arousal level on a scale from 1 (very calm/soothing) to 9 (very intense/energetic).
Low arousal: Relaxing, calm, sleepy, or soothing music
High arousal: Exciting, intense, energetic, or stimulating music
Return a single number between 1 and 9 representing the perceived arousal level.", 
"input": "<|SOA|><AUDIO><|EOA|>", 
"output": "0.4645976536872577", 
"uuid": "", 
"split": ["train"], 
"task_type": {"major": ["regression"], 
"minor": ["emotion_regression"]}, 
"domain": "music", 
"audio_path": ["data/EMO/emomusic/wav/0020.wav"], 
"audio_start": 0.0, 
"audio_end": 45.0, 
"source": "EMO", 
"other": {"tag": "null"}}
\end{lstlisting}
\begin{lstlisting}
{"instruction": "Valence refers to the emotional quality or pleasantness of the emotion experienced, whether the emotion is positive or negative.
You are given a piece of music. Estimate its valence level on a scale from 1 (very sad/melancholic) to 9 (very happy/cheerful).
Positive valence : Happy, joyful, content.
Negative valence : Sad, angry, anxious.
Return a single number between 1 and 9 representing the perceived Valence level.",
"input": "<|SOA|><AUDIO><|EOA|>", 
"output": "2.268336892705501", 
"uuid": "", 
"split": ["train"], 
"task_type":{"major": ["regression"], 
"minor": ["emotion_regression"]}, 
"domain": "music", 
"audio_path": ["data/EMO/emomusic/wav/0025.wav"], 
"audio_start": 0.0, 
"audio_end": 45.0, 
"source": "EMO", 
"other": {"tag": "null"}}
\end{lstlisting}
\begin{lstlisting} 
{"instruction": "Please provide the tags of given music.", "input": "<|SOA|><AUDIO><|EOA|>", "output": "classical, opera, singing, quiet, woman, voice, female voice, choral", "uuid": "", "split": ["train"],
"task_type": {"major": ["multi_label"], "minor": ["music_tagging"]}, 
"domain": "music",
"audio_path": ["data/MTT/mp3/9/american_bach_soloists-heinrich_schutz__musicalische_exequien-01-musicalische_exequien_swv_279_teil_i_concert_in_form_einer_teutschen_begrabnismissa-349-378.mp3"], 
"audio_start": 0.0, 
"audio_end": 30.0, 
"source": "The MagnaTagATune Dataset", 
"other": {"tag": "null"}}\end{lstlisting}
\begin{lstlisting} 
{"instruction": "You are given an audio clip of a music piece. Identify the genre by selecting exactly one option from the following list: 
Genres: blues, classical, country, disco, hip-hop, jazz, metal, pop, reggae, rock.
Return only the name of the predicted genre. No explanation.", 
"input": "<|SOA|><AUDIO><|EOA|>",
"output": "blues", 
"uuid": "", 
"split": ["train"],
"task_type": {"major": ["multi_class"],
"minor": ["genre_classification"]}, 
"domain": "music",
"audio_path": ["data/GTZAN/Data/genres_original/blues/blues.00034.wav"], 
"audio_start": 0.0, 
"audio_end": 30.0, 
"source": "GTZAN Dataset", 
"other": {"tag": "null"}}
\end{lstlisting}
\begin{lstlisting} 
{"instruction": "You are given an audio clip of a vocal performance.
Identify the singing technique used by selecting exactly one option from the list below:
Singing Techniques: belt, breathy, inhaled, lip_trill, spoken, straight, trill, trillo, vibrato, vocal_fry.
Return only the name of the predicted technique. No explanation", 
"input": "<|SOA|><AUDIO><|EOA|>", 
"output": "belt", 
"uuid": "", 
"split": ["train"], 
"task_type":{"major": ["multi-class"], 
"minor": ["singing_technique_classification"]}, 
"domain": "music", 
"audio_path": ["data/VocalSet/audio/belt/f1_scales_belt_o_01.wav"], "audio_start": 0.0, "audio_end": -1, 
"source": "VocalSet",
"other": {"tag": "null"}}
\end{lstlisting}
\begin{lstlisting}
{"instruction": "Identify the pitch of the given audio. You must choose exactly one pitch from the given range.\n\n
Pitches are represented using the MIDI standard, where each pitch is assigned an integer value between 0 (lowest) and 127 (highest). In this dataset, the lowest pitch is 9 (1 octave below A_0, 13.75Hz), and the highest pitch is 119 (B 8, 7902.13Hz). Be sure to convert a musical note to its MIDI number:\n
Middle C (C4) = 60\nA440 (A4) = 69\n
One octave higher = +12 MIDI numbers\nOne octave lower = -12 MIDI numbers\n\n
Your response should contain only one MIDI pitch number.\n\n
Example 1, when the note is A1: 33\n
Example 2, when the note is F5: 77\n
Example 3, when the note is A4: 69", 
"input": "<|SOA|><AUDIO><|EOA|>", 
"output": 84, "uuid": "", "split": ["test"], "task_type": \begin{lstlisting} {"major": ["multi-class"], 
"minor": ["pitch_classification"]}, 
"domain": "music", 
"audio_path": ["data/NSynth/nsynth-test/audio/bass_synthetic_068-084-050.wav"], "audio_start": 0.0, 
"audio_end": -1, 
"source": "Nsynth", 
"other": {"tag": "null"}}\end{lstlisting}
\begin{lstlisting}
{"instruction": "Please transcribe the lyrics of the given song.", 
"input": "<|SOA|><AUDIO><|EOA|>", 
"output": "DOESN'T TEAR YOU APART ANYMORE", 
"uuid": "", 
"split": ["test"], 
"task_type": {"major": ["transcription"], 
"minor": ["lyrics_transcription"]}, 
"domain": "music", 
"audio_path": ["data/DSing/sing_300x30x2/GB/GBVocals/225741816_226995-475216816_227615877-GB-F-475215813.m4a"], 
"audio_start": 277.867, 
"audio_end": 283.778, 
"source": "DSing", 
"other":{"tag": "null"}}
\end{lstlisting}
\begin{lstlisting} 
{"instruction": "Identify and list the timestamps of all downbeats in this audio track. Use the format of `0.0s,0.54s,1.0ss, ...`", 
"input": "<|SOA|><AUDIO><|EOA|>", 
"output": "1.61s,3.42s,5.25s,7.04s,8.85s,10.65s,12.47s,14.27s,16.08s,17.88s,19.68s,21.5s,23.32s,25.12s,26.94s,28.74s", 
"uuid": "", 
"split": ["test"],
"task_type": {"major": ["seq_multi-class"], 
"minor": ["downbeat_tracking"]}, "audio_path": ["data/ballroom/BallroomData/Rumba-American/Albums-AnaBelen_Veneo-13.wav"], "domain": "music", 
"audio_start": 0.0, 
"audio_end": 30.0, 
"source": "ballroom", 
"other": {"tag": "null"}}\
\end{lstlisting}

\begin{lstlisting} 
{"instruction": "Detect the timestep occurrences of Guzheng (Chinese zither) playing techniques in the given audio. The possible techniques include: Vibrato, Point Note, Upward Portamento, Downward Portamento, Plucks, Glissando, and Tremolo.\n\n
The output format should be a Python string representation of a list containing tuples of (start time second, end time second, technique). If no technique is detected, return [('start_time', 'end_time', 'No Tech')].\n\n
Example 1:\n
\"[('5.5035', '6.0724', 'Upward Portamento'), ('7.0708', '8.0809', 'Upward Portamento'), ('9.6947', '10.0', 'Upward Portamento')]\"\n
Example 2:\n
\"[('0.0', '10.0', 'No Tech')]\" ", 
"input": "<|SOA|><AUDIO><|EOA|>", 
"output": "[('9.7591', '10.4321', 'Upward Portamento')]", 
"uuid": "", 
"split": ["train"], 
"task_type": {"major": ["seq_multi-class"], 
"minor": ["technique_detection"]}, 
"domain": "music", 
"audio_path": ["data/Guzheng/Guzheng_Tech99/data/audio/train/41.flac"], 
"audio_start": 0.0, 
"audio_end": 10.0, 
"source": "Guzheng Tech99", 
"other":  {"tag": "null"}}\end{lstlisting}

\subsection{Comparison with Previous Benchmarks}

\begin{figure}[htb]
  \centering
  \includegraphics[alt={ISMIR 2025 template example image},width=0.85\linewidth]{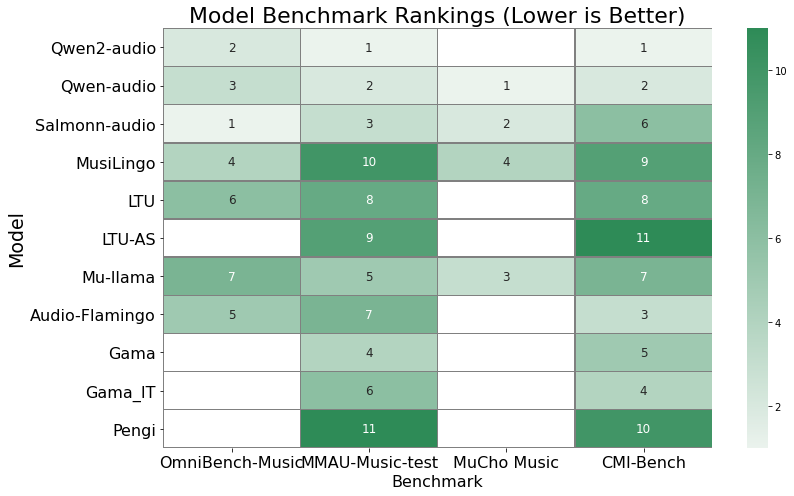}
  \caption{Comparison of model rankings across four benchmarks: OmniBench-Music, MMAU-Music-test, MuCho Music, and our proposed CMI-Bench. While absolute scores vary across benchmarks, the relative ordering of models remains largely consistent, suggesting stable performance trends across different evaluation protocols.}
  \label{fig:example}\label{fig:four-bench}
\end{figure}
For each task and each dataset in CMI-bench, we average all evaluation metrics (e.g., average 4 metrics for captioning on SDD and MusicCaps respectively) and rank all model performances from 0 to 10. For each model, we average the rank of each model among all tasks to get the universal rank on CMI-benchmark.

Our evaluation across four benchmarks—OmniBench-Music, MMAU-Music-test, MuCho Music, and our proposed benchmark—reveals consistent performance trends among the tested LLMs for music-related tasks. As shown in the figure \ref{fig:four-bench}, models like Qwen-audio and Salmonn-audio exhibit robust rankings across all benchmarks, suggesting their adaptability to diverse music understanding challenges. 
The similarity in model ordering across benchmarks underscores a degree of stability in their relative capabilities, yet the nuanced differences
suggest that task-specific design or training data may play a critical role. These findings invite further exploration into how LLMs balance generalizability and specialization in MIR tasks.

\begin{figure*}[h]
  \centering
  \includegraphics[alt={ISMIR 2025 template example image},width=0.97\linewidth]{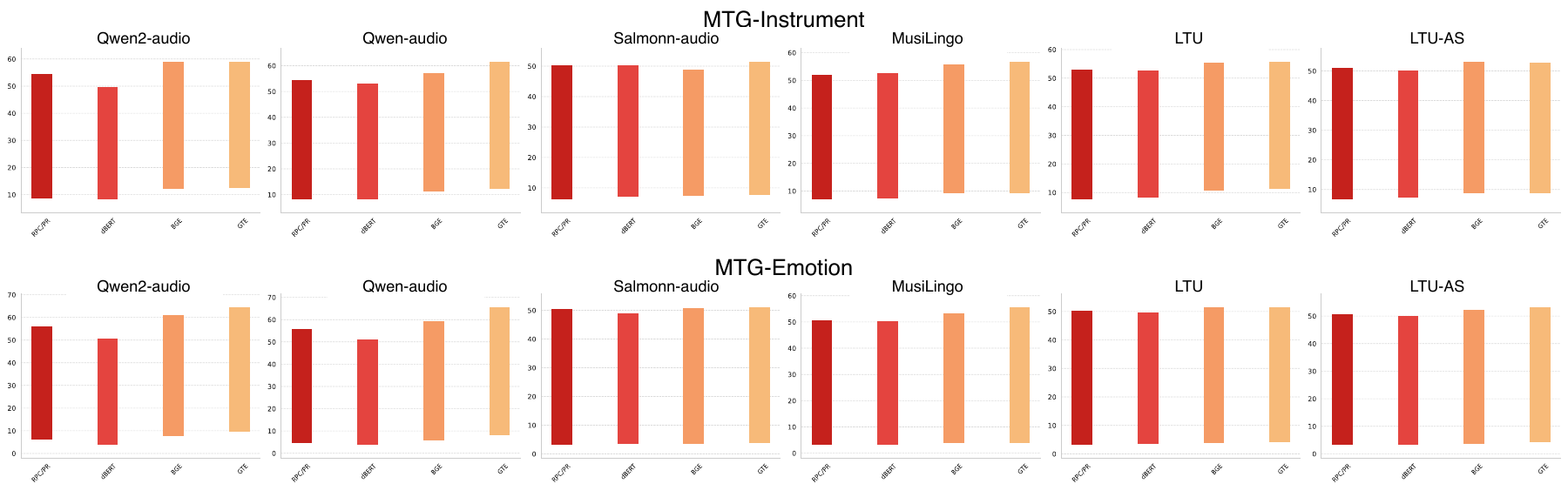}
  \caption{Comparison of multi-label evaluation methods on MTG-Emotion and MTG-Instrument tasks. "Accurate" denotes strict matching, while BERTScore, BGE, and GTE-Qwen represent embedding-based approaches..}
  \label{fig:example}
\end{figure*}

\subsection{Embedding-Based Evaluation for Multi-Label Tagging}

In multi-label tagging tasks such as emotion and instrument classification, model outputs often include free-form text that may not exactly match the ground-truth tags. To address this, we compare several retrieval-based approaches for computing ROC-AUC and PR-AUC: (1) exact match (“Accurate”), (2) semantic similarity using BERTScore (roberta-large), (3) BGE encoder, and (4) GTE-Qwen embedding model. As shown in Table X, exact match yields the lowest scores, failing to capture near-synonymous predictions (e.g., "electric guitar" vs. "guitar"). While BERTScore offers modest improvements, it is not optimized for retrieval and underperforms in multi-label scenarios. BGE, a compact 330M model fine-tuned for text embedding and retrieval, consistently delivers strong results while remaining computationally efficient. GTE-Qwen, a larger 7B model with strong retrieval capabilities, achieves the best scores overall but is significantly more expensive to run. Based on this analysis, we adopt BGE for all benchmark evaluations, striking a balance between retrieval quality and evaluation efficiency.

\subsection{Case Study of Qwen-audio Output}
\subsubsection{Beat Tracking}
\begin{lstlisting} 
 {
        "question": "Identify and list the timestamps of all downbeats in this audio track. Use the format of `0.0s,0.54s,1.0ss, ...`",
        "response": "The timestamps of all downbeats in this audio track are `0.0s,0.54s,1.0ss, 1.62s, 2.16s, 2.70s, 3.24s, 3.78s, 4.32s, 4.86s, 5.40s, 5.94s, 6.48s, 7.02s, 7.56s, 8.10s, 8.64s, 9.18s, 9.72s, 10.26s, 10.80s, 11.34s, 11.88s, 12.42s, 12.96s, 13.50s, 14.04s, 14.58s, 15.12s, 15.66s, 16.20s, 16.74s, 17.28s, 17.82s, 18.36s, 18.90s, 19.44s, 19.98s, 20.52s, 21.06s, 21.60s, 22.14s, 22.68s, 23.22s, 23.76s, 24.30s, 24.84s, 25.38s, 26.02s, 26.56s, 27.10s, 27.64s, 28.18s, 28.72s, 29.26s, 29.80s, 30.34s, 30.88s, 31.42s, 31.96s, 32.50s, 33.04s, 33.58s, 34.12s, 34.66s, 3",
        "correct_answer": "1.61s,3.42s,5.25s,7.04s,8.85s,10.65s,12.47s,14.27s,16.08s,17.88s,19.68s,21.5s,23.32s,25.12s,26.94s,28.74s",
        "audioid": "../testdata/ballroom/BallroomData/Rumba-American/Albums-AnaBelen_Veneo-13.wav",
        "other": ""
    },
\end{lstlisting}
The downbeat tracking output of Qwen-audio on ballroom dataset, though it provides the best results among all models, are simply copying the input example always.
\subsubsection{Lyrics Transcription}
\begin{lstlisting} 
{
        "question": "Please transcribe the lyrics of the given song.",
        "response": "The lyrics of the given song are:\"You make me feel alive\".",
        "correct_answer": "YOU MADE ME FEEL AS THOUGH",
        "audioid": "../testdata/DSing/sing_300x30x2/GB/GBVocals/3770008_3770008-1036593237_1553518113-GB-F-933206912.m4a",
        "other": ""
    },
\end{lstlisting}
The results of Qwen-audio do not have hallucination on song description, music captions, lyrics context or emoji that are typically produced by other models, but we can still observe significant errors in the upper cases. 

Besides, the model can generate the same word in the lyrics repeatedly without a stop when the singer sing a word twice or three times, as illustrated by the following case. If we exclude such cases in the testset, the WER performance of Qwen-audio model will decrease from 115.7(\%) to 50(\%), which is still significantly worse than the SOTA performance~13(\%), but much more reasonable.

\begin{lstlisting}
{High WER: 167.66666666666666,
    prediction: The lyrics of the given song are:"Go go go go go go go go go go go go go go go go go go go go go go go go go go go go go go go go go go go go go go go go go go go go go go go go go go go go go go go go go go go go go go go go go go go go go go go go go go go go go go go go go go go go go go go go go go go go go go go go go go go go go go go go go go go go go go go go go go go go go go go go go go go go go go go go go go go go go go go go go go go go go go go go go go go go go go go go go go go go go go go go go go go go go go go go go go go go go go go go go go go go go go go go go go go go go go go go go go go go go go go go go go go go go go go go go go go go go go go go go go go go go go go go go go go go go go go go go go go go go go go go go go go go go go go go go go go go go go go go go go go go go go go go go go go go go go go go go go go go go go go go go go go go go go go go go go go go go go go go go go go go go go go go go go go go go go go go go go go go go go go go go go go go go go go go go go go go go go go go go go go go go go go go go go go go go go go go go go go go go go go go go go go go go go go go go go go go go go go go go go go go go go go go go go go go go go go go go go go go go go go go go go go go go go go go go go go go go go go go go go go go go go go go go go go go go go go go go go go go go go go go go go go go go go go go go go go go go go go go go go go go go go go go go go go go go go go go go go go go go go go go go go go go go go go go go go go go go go go go go go go go go go go,
    reference: DON'T DON'T DON'T},
{WER: 62.0,
prediction: The lyrics of the given song are:"Cause you knew that I knew that I knew that I knew that I knew that I knew that I knew that I knew that I knew that I knew that I knew that I knew that I knew that I knew that I knew that I knew that I knew that I knew that I knew that I knew that I knew that I knew that I knew that I knew that I knew that I knew that I knew that I knew that I knew that I knew that I knew that I knew that I knew that I knew that I knew that I knew that I knew that I knew that I knew that I knew that I knew that I knew that I knew that I knew that I knew that I knew that I knew that I knew that I knew that I knew that I knew that I knew that I knew that I knew that I knew that I knew that I knew that I knew that I knew that I knew that I knew that I knew that I knew that I knew that I knew that I knew that I knew that I knew that I knew that I knew that I knew that I knew that I knew that I knew that I knew that I knew that I knew that I knew that I knew that I knew that I knew that I knew that I knew that I knew that I knew that I knew that I knew that I knew that I knew that I knew that I knew that I knew that I knew that I knew that I knew that I knew that I knew that I knew that I knew that I knew that I knew that I knew that I knew that I knew that I knew that I knew that I knew that I knew that I knew that I knew that I knew that I knew that I knew that I knew that I knew that I knew that I knew that I knew that I knew that I knew that I knew that I knew that I knew that I knew that I knew that I knew that I knew that I knew that I knew that I knew that I knew that I knew that I knew that I knew that I knew that I knew that I knew that I knew that I knew that I knew that I knew that I knew that I knew that I knew that I knew that I knew that I knew that I knew that I knew that I knew that I knew that I knew that I knew that I knew that I knew that I knew that I knew that I knew that I knew that I knew that I knew that I knew that I knew that I knew that I knew that I knew that I knew that I,
reference: 'CAUSE YOU KNEW THAT I KNEW THAT I
},
{WER: 2.25,
prediction: The lyrics of the given song are:"Black black black black black black Black black black black black black".
reference: BLACK BLACK OH BLACK}
\end{lstlisting}
\subsubsection{EMO}
\begin{lstlisting}
    {
        "question": "Arousal refers to the level of physiological or psychological activation a piece of music induces in a listener. It reflects how calming or energizing the music feels.\nYou are given a piece of music. Estimate its arousal level on a scale from 1 (very calm/soothing) to 9 (very intense/energetic).\nLow arousal: Relaxing, calm, sleepy, or soothing music\nHigh arousal: Exciting, intense, energetic, or stimulating music\nReturn a single number between 1 and 9 representing the perceived arousal level.",
        "response": "Based on the audio, I would estimate a arousal level of 6.",
        "correct_answer": "1.9075890449018524",
        "audioid": "../testdata/EMO/emomusic/wav/0962.wav",
        "other": ""
    },
    {
        "question": "Arousal refers to the level of physiological or psychological activation a piece of music induces in a listener. It reflects how calming or energizing the music feels.\nYou are given a piece of music. Estimate its arousal level on a scale from 1 (very calm/soothing) to 9 (very intense/energetic).\nLow arousal: Relaxing, calm, sleepy, or soothing music\nHigh arousal: Exciting, intense, energetic, or stimulating music\nReturn a single number between 1 and 9 representing the perceived arousal level.",
        "response": "Based on the audio, it sounds like the music is medium to high arousal.",
        "correct_answer": "1.330392488416014",
        "audioid": "../testdata/EMO/emomusic/wav/0970.wav",
        "other": ""
    },
\end{lstlisting}
Qwen-audio response on EMO sometimes can give a reasonable result, but not always. Besides %
it may disobey the instruction commonly.

\subsubsection{GS-Key}
\begin{lstlisting}
    {
        "question": "You are given an audio clip of a music piece.\nEstimate the musical key of the audio by selecting exactly one key from the following list:\n\nC major, Db major, D major, Eb major, E major, F major, Gb major, G major, Ab major, A major, Bb major, B major,\nC minor, Db minor, D minor, Eb minor, E minor, F minor, Gb minor, G minor, Ab minor, A minor, Bb minor, B minor.\n\nReturn only the name of the predicted key. No explanation.\n",
        "response": "The musical key of the audio is C major.",
        "correct_answer": "D major",
        "audioid": "../testdata/GS-key/giantsteps_clips/wav/0010089-3.wav",
        "other": ""
    },
    {
        "question": "You are given an audio clip of a music piece.\nEstimate the musical key of the audio by selecting exactly one key from the following list:\n\nC major, Db major, D major, Eb major, E major, F major, Gb major, G major, Ab major, A major, Bb major, B major,\nC minor, Db minor, D minor, Eb minor, E minor, F minor, Gb minor, G minor, Ab minor, A minor, Bb minor, B minor.\n\nReturn only the name of the predicted key. No explanation.\n",
        "response": "The musical key of the audio is A major.",
        "correct_answer": "F minor",
        "audioid": "../testdata/GS-key/giantsteps_clips/wav/0061415-0.wav",
        "other": ""
    },
\end{lstlisting}
Qwen-audio can follow the instruction of Key detection, but its ZERO-SHOT performance does not surpass the random guess.

\subsubsection{Genre Classification}
\begin{lstlisting}
    {
        "question": "You are given an audio clip of a music piece. Identify the genre by selecting exactly one option from the following list: \nGenres: blues, classical, country, disco, hip-hop, jazz, metal, pop, reggae, rock.\nReturn only the name of the predicted genre. No explanation.\n",
        "response": "The predicted genre is blues.",
        "correct_answer": "blues",
        "audioid": "../testdata/GTZAN/Data/genres_original/blues/blues.00027.wav",
        "other": ""
    },
    {
        "question": "You are given an audio clip of a music piece. Identify the genre by selecting exactly one option from the following list: \nGenres: blues, classical, country, disco, hip-hop, jazz, metal, pop, reggae, rock.\nReturn only the name of the predicted genre. No explanation.\n",
        "response": "The predicted genre is blues.",
        "correct_answer": "blues",
        "audioid": "../testdata/GTZAN/Data/genres_original/blues/blues.00028.wav",
        "other": ""
    },
\end{lstlisting}
Qwen-audio can provide reasonable results on clip-level multi-class such as genre classification.

\subsubsection{Pitch Recognition}
\begin{lstlisting} 
    {
        "question": "Identify the pitch of the given audio. You must choose exactly one pitch from the given range.\n\nPitches are represented using the MIDI standard, where each pitch is assigned an integer value between 0 (lowest) and 127 (highest). In this dataset, the lowest pitch is 9 (1 octave below A_0, 13.75Hz), and the highest pitch is 119 (B 8, 7902.13Hz). Be sure to convert a musical note to its MIDI number:\nMiddle C (C4) = 60\nA440 (A4) = 69\nOne octave higher = +12 MIDI numbers\nOne octave lower = -12 MIDI numbers\n\nYour response should contain only one MIDI pitch number.\n\nExample 1, when the note is A1: 33\nExample 2, when the note is F5: 77\nExample 3, when the note is A4: 69",
        "response": "The given audio is the recording of a voice message saying, \"This is the message\". The frequency of the voice message is around 135 Hz.",
        "correct_answer": 49,
        "audioid": "../testdata/NSynth/nsynth-test/audio/bass_synthetic_068-049-025.wav",
        "other": ""
    },
\end{lstlisting}
135Hz is C\#3 minus 45 cents, the MIDI number result is 49, but it does not follow the instruction format.

\subsubsection{Vocal Technique Classification}
\begin{lstlisting}
        {
        "question": "You are given an audio clip of a vocal performance.\nIdentify the singing technique used by selecting exactly one option from the list below:\nSinging Techniques: belt, breathy, inhaled, lip_trill, spoken, straight, trill, trillo, vibrato, vocal_fry.\nReturn only the name of the predicted technique. No explanation.",
        "response": "The singing technique used is \"belt\".",
        "correct_answer": "belt",
        "audioid": "../testdata/VocalSet/audio/belt/f2_arpeggios_belt_o_00.wav",
        "other": ""
    },
    {
        "question": "You are given an audio clip of a vocal performance.\nIdentify the singing technique used by selecting exactly one option from the list below:\nSinging Techniques: belt, breathy, inhaled, lip_trill, spoken, straight, trill, trillo, vibrato, vocal_fry.\nReturn only the name of the predicted technique. No explanation.",
        "response": "The singing technique used is \"vocal fry\".",
        "correct_answer": "belt",
        "audioid": "../testdata/VocalSet/audio/belt/f2_arpeggios_belt_u_00.wav",
        "other": ""
    },
    {
        "question": "You are given an audio clip of a vocal performance.\nIdentify the singing technique used by selecting exactly one option from the list below:\nSinging Techniques: belt, breathy, inhaled, lip_trill, spoken, straight, trill, trillo, vibrato, vocal_fry.\nReturn only the name of the predicted technique. No explanation.",
        "response": "The singing technique used is \"vibrato\".",
        "correct_answer": "belt",
        "audioid": "../testdata/VocalSet/audio/belt/f2_scales_belt_a_00.wav",
        "other": ""
    },
\end{lstlisting}
On Vocal technique classification, Qwen-audio can always follow the instruction, but provides the wrong or biased answer. %

\subsubsection{Other tasks}
\begin{lstlisting} 
{
        "question": "Detect the timestep occurrences of Guzheng (Chinese zither) playing techniques in the given audio. The possible techniques include: Vibrato, Point Note, Upward Portamento, Downward Portamento, Plucks, Glissando, and Tremolo.\n\nThe output format should be a Python string representation of a list containing tuples of (start time second, end time second, technique). If no technique is detected, return [('start_time', 'end_time', 'No Tech')].\n\nExample 1:\n\"[('5.5035', '6.0724', 'Upward Portamento'), ('7.0708', '8.0809', 'Upward Portamento'), ('9.6947', '10.0', 'Upward Portamento')]\"\nExample 2:\n\"[('0.0', '10.0', 'No Tech')]\" ",
        "response": "The given audio contains a guzheng playing a melody. Using a voice recognition model, the lyrics are \"\u5c71\u5916\u9752\u5c71\u697c\u5916\u697c\". The song genre is Chinese Traditional.",
        "correct_answer": "[('140.3988', '141.6056', 'Vibrato'), ('142.8589', '144.0425', 'Vibrato'), ('145.7715', '146.3285', 'Vibrato'), ('148.1736', '148.7770', 'Vibrato'), ('149.9955', '151.2023', 'Vibrato')]",
        "audioid": "../testdata/Guzheng/Guzheng_Tech99/data/audio/test/94.flac",
        "other": ""
    },
    {
        "question": "Detect the timestep occurrences of Guzheng (Chinese zither) playing techniques in the given audio. The possible techniques include: Vibrato, Point Note, Upward Portamento, Downward Portamento, Plucks, Glissando, and Tremolo.\n\nThe output format should be a Python string representation of a list containing tuples of (start time second, end time second, technique). If no technique is detected, return [('start_time', 'end_time', 'No Tech')].\n\nExample 1:\n\"[('5.5035', '6.0724', 'Upward Portamento'), ('7.0708', '8.0809', 'Upward Portamento'), ('9.6947', '10.0', 'Upward Portamento')]\"\nExample 2:\n\"[('0.0', '10.0', 'No Tech')]\" ",
        "response": "The given audio contains a guzheng playing a melody. Using a music recognition model, the following steps can be taken to extract the techniques used in the melody:\n\n  1. Apply music recognition to identify the melody.\n  2. Extract the notes played in the melody.\n  3. Identify the techniques used in the melody based on the notes played.\n\nThe following is the Python code to extract the techniques used in the given audio:\n```python\nimport music21\n\n# Load the audio file\naudio = music21.midi.readAudioFile(\"audio_file.mid\")\n\n# Create a stream from the audio file\nstream = audio.createStream()\n\n# Get the note values and durations from the stream\nnotes = stream.getNoteValues()\n\n# Identify the techniques used in the melody\ntechniques = []\nfor note in notes:\n    if note.velocity < 40:\n        techniques.append(('Vibrato', note.start, note.end))\n    elif note.velocity == 40:\n        techniques.append(('Point Note', note.start, note.end))\n    elif note.velocity > 40:\n        techniques.append(('Upward Portamento', note.start, note.end))\n        techniques.append(('Downward Portamento', note.start, note.end))\n    elif note.velocity < -40:\n        techniques.append(('Plucks', note.start, note.end))\n    elif note.velocity == -40:\n        techniques.append(('Glissando', note.start, note.end))\n    elif note.velocity > -40:\n        techniques.append(('Tremolo', note.start, note.end))\n\n# Print the techniques detected in the melody\nprint(techniques)\n```\n\nThe output of this code will be a list of tuples containing the start time, end time, and technique used for each note in the melody. For example:\n```python\n[('0.0', '0.3437', 'Vibrato'), ('0.3437', '0.6474', 'Vibrato'), ('0.6474', '0.9511', 'Vibrato'), ('0.9511', '1.2548', 'Vibrato'), ('1.2548', '1.5585', 'Vibrato'), ('1.5585', '1.8622', 'Vibrato'), ('1.8622', '2.165",
        "correct_answer": "[('154.1962', '154.7532', 'Vibrato')]",
        "audioid": "../testdata/Guzheng/Guzheng_Tech99/data/audio/test/94.flac",
        "other": ""
    },
    {
        "question": "Detect the timestep occurrences of Guzheng (Chinese zither) playing techniques in the given audio. The possible techniques include: Vibrato, Point Note, Upward Portamento, Downward Portamento, Plucks, Glissando, and Tremolo.\n\nThe output format should be a Python string representation of a list containing tuples of (start time second, end time second, technique). If no technique is detected, return [('start_time', 'end_time', 'No Tech')].\n\nExample 1:\n\"[('5.5035', '6.0724', 'Upward Portamento'), ('7.0708', '8.0809', 'Upward Portamento'), ('9.6947', '10.0', 'Upward Portamento')]\"\nExample 2:\n\"[('0.0', '10.0', 'No Tech')]\" ",
        "response": "The Guzheng (Chinese zither) is playing techniques such as Vibrato, Point Note, Upward Portamento, Downward Portamento, Plucks, Glissando, and Tremolo in the given audio. Here is the list of detected techniques with their corresponding start and end times in seconds: [(5.5035, 6.0724, 'Upward Portamento'), (7.0708, 8.0809, 'Upward Portamento'), (9.6947, 10.0, 'Upward Portamento')]",
        "correct_answer": "[('0.0', '10.0', 'No Tech')]",
        "audioid": "../testdata/Guzheng/Guzheng_Tech99/data/audio/test/94.flac",
        "other": ""
    }
\end{lstlisting}
Qwen-audio cannot follow the instruction not only on task description but format requirement on instrument performance techniques detection. Not to mention the under-represented instrument of world music.
\begin{lstlisting} 
{
        "question": "Please provide the emotion tag(s) of given audio.",
        "response": "The emotion tag(s) of given audio are: relax, calm, chill, peaceful, spa, easy listening, lounge, sleep, meditation, yoga, relax, ambient, downtempo, background, instrumental, ambient, easy listening, lounge, sleep, meditation, yoga, relax, calm, chill, peaceful, spa, easy listening, lounge, sleep, meditation, yoga, relax, ambient, downtempo, background, instrumental.",
        "correct_answer": "meditative",
        "audioid": "../testdata/MTG/audio-low/83/5083.low.mp3",
        "other": ""
    },
    {
        "question": "Please provide the emotion tag(s) of given audio.",
        "response": "The audio is of a jazz song.",
        "correct_answer": "cool",
        "audioid": "../testdata/MTG/audio-low/84/5084.low.mp3",
        "other": ""
    },
\end{lstlisting}
\begin{lstlisting} 
{
        "question": "Please provide the genre tag(s) of given audio.",
        "response": "Music genre: Ambient, Chill-out, Lounge, New Age, Electronic, IDM, Ambient House, Chill-out, Lounge, New Age, Electronic, IDM, Ambient, Chill-out, Lounge, New Age, Electronic, IDM, Ambient, Chill-out, Lounge, New Age, Electronic, IDM, Ambient, Chill-out, Lounge, New Age, Electronic, IDM, Ambient, Chill-out, Lounge, New Age, Electronic, IDM, Ambient, Chill-out, Lounge, New Age, Electronic, IDM, Ambient, Chill-out, Lounge, New Age, Electronic, IDM, Ambient, Chill-out, Lounge, New Age, Electronic, IDM, Ambient, Chill-out, Lounge, New Age, Electronic, IDM, Ambient, Chill-out, Lounge, New Age, Electronic, IDM, Ambient, Chill-out, Lounge, New Age, Electronic, IDM, Ambient, Chill-out, Lounge, New Age, Electronic, IDM, Ambient, Chill-out, Lounge, New Age, Electronic, IDM, Ambient, Chill-out, Lounge, New Age, Electronic, IDM, Ambient, Chill-out, Lounge, New Age, Electronic, IDM, Ambient, Chill-out, Lounge, New Age, Electronic, IDM, Ambient, Chill-out, Lounge, New Age, Electronic, IDM, Ambient, Chill-out, Lounge, New Age, Electronic, IDM, Ambient, Chill-out, Lounge, New Age, Electronic, IDM, Ambient, Chill-out, Lounge, New Age, Electronic, IDM, Ambient, Chill-out, Lounge, New Age, Electronic, IDM, Ambient, Chill-out, Lounge, New Age, Electronic, IDM, Ambient, Chill-out, Lounge, New Age, Electronic, IDM, Ambient, Chill-out, Lounge, New Age, Electronic, IDM, Ambient, Chill-out, Lounge, New Age, Electronic, IDM, Ambient, Chill-out, Lounge, New Age, Electronic, IDM, Ambient, Chill-out, Lounge, New Age, Electronic, IDM, Ambient, Chill-out, Lounge, New Age, Electronic, IDM, Ambient, Chill-out, Lounge, New Age, Electronic, IDM, Ambient, Chill-out, Lounge, New Age, Electronic, IDM, Ambient, Chill-out, Lounge, New Age, Electronic, IDM, Ambient, Chill-out, Lounge, New Age, Electronic, IDM, Ambient, Chill-out, Lounge, New Age, Electronic, IDM, Ambient, Chill-out, Lounge, New Age, Electronic, IDM, Ambient, Chill-out, Lounge, New Age, Electronic, IDM, Ambient, Chill",
        "correct_answer": "ambient, easylistening, orchestral",
        "audioid": "../testdata/MTG/audio-low/50/1420850.low.mp3",
        "other": ""
    },
    {
        "question": "Please provide the genre tag(s) of given audio.",
        "response": "Jazz Lounge 1",
        "correct_answer": "downtempo, electronic, jazzfusion",
        "audioid": "../testdata/MTG/audio-low/99/1420899.low.mp3",
        "other": ""
    },
    {
        "question": "Please provide the genre tag(s) of given audio.",
        "response": "The genre tag(s) of given audio are electronic, experimental, electroacoustic, instrumental.",
        "correct_answer": "electronic, house, techno",
        "audioid": "../testdata/MTG/audio-low/14/1421614.low.mp3",
        "other": ""
    }
\end{lstlisting}
\begin{lstlisting} 
{
        "question": "Please provide the instrument tag(s) of given audio.",
        "response": "The instrument tag(s) of given audio are nylon string guitar.",
        "correct_answer": "guitar",
        "audioid": "../testdata/MTG/audio-low/88/4888.low.mp3",
        "other": ""
    },
    {
        "question": "Please provide the instrument tag(s) of given audio.",
        "response": "The instrument tag(s) of given audio are: drum, piano, bass.",
        "correct_answer": "piano",
        "audioid": "../testdata/MTG/audio-low/90/4890.low.mp3",
        "other": ""
    },
\end{lstlisting}
Qwen-audio sometimes mistake the emotion instruction for the genre and instruments. While the results of genre and instrument labelling are relatively reasonable.

\begin{lstlisting} 
{
        "question": "Please provide the description of given song.",
        "response": "This is a song whose genre is Electronic, and the lyrics are \"Butterflied wings you'll find\".",
        "correct_answer": "Electronic music that has a constant melody throughout with accompanying instruments used to supplement the melody which can be heard in possibly a casual setting",
        "audioid": "../testdata/SDD/audio/audio/34/1004034.2min.mp3",
        "other": ""
    },
    {
        "question": "Please provide the description of given song.",
        "response": "Solo electric guitar",
        "correct_answer": "acoustic guitar solo track with consistent rhythm and repeating progression, suitable for a relaxing afternoon tea",
        "audioid": "../testdata/SDD/audio/audio/74/1007274.2min.mp3",
        "other": ""
    },
    {
        "question": "Please provide the description of given song.",
        "response": "This is a song whose genre is Pop, and the lyrics are \"See the magic circle that surrounds my heart We need to touch it when he bites my hand There's a mystical drawing somewhere in my mind Something about love something higher than you do\".",
        "correct_answer": "A rock song with a country vibe, it has male vocals, drums, and acoustic guitar. It is an upbeat song",
        "audioid": "../testdata/SDD/audio/audio/71/1009671.2min.mp3",
        "other": ""
    },
\end{lstlisting}
Qwen-audio has reasonable results on caption generation. 

\subsection{PCA Components of CMI-Benchmark}
We regard the 34 metrics of CMI-benchmark as the 34 dimensions of LLM MIR capabilities, and we evaluate the PCA results of the 11 observation of models. The top 4 PCA components are as follows:

\begin{figure}[htbp]
    \centering

    \subfigure[PCA component 1]{
        \includegraphics[width=0.9\linewidth]{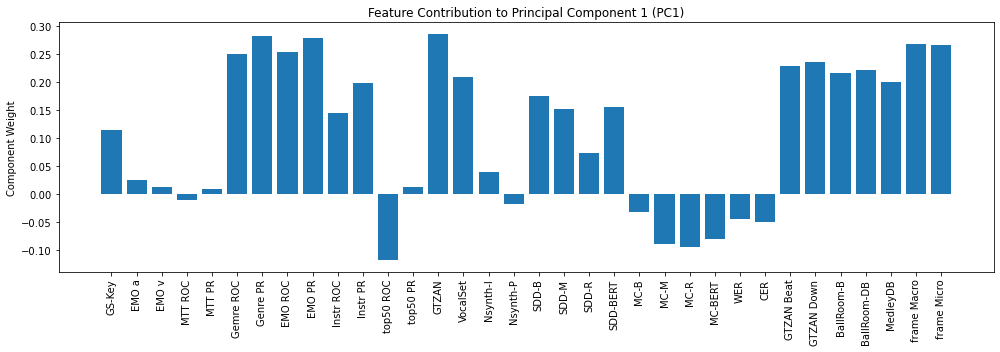}
        \label{fig:pca1}
    }\\[1ex]

    \subfigure[PCA component 2]{
        \includegraphics[width=0.9\linewidth]{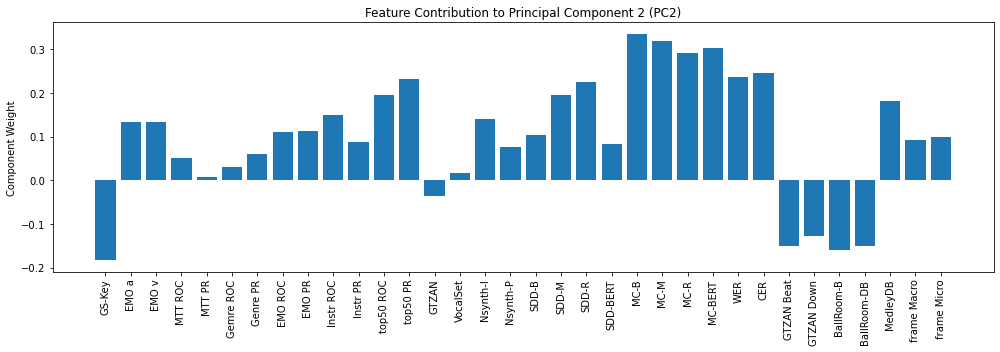}
        \label{fig:pca2}
    }\\[1ex]

    \subfigure[PCA component 3]{
        \includegraphics[width=0.9\linewidth]{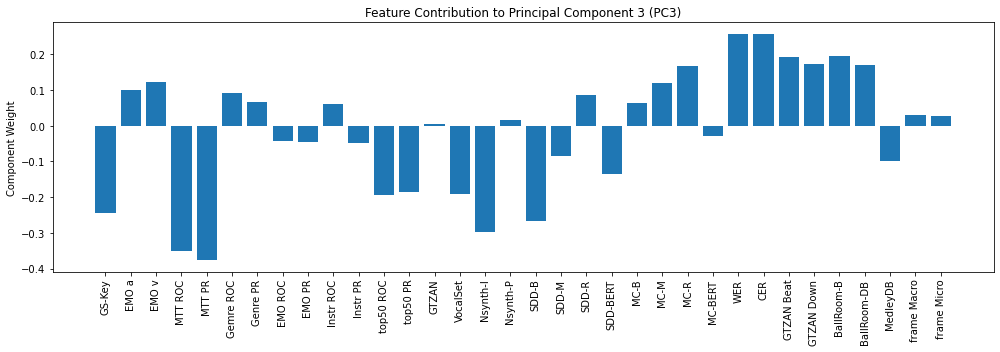}
        \label{fig:pca3}
    }\\[1ex]

    \subfigure[PCA component 4]{
        \includegraphics[width=0.9\linewidth]{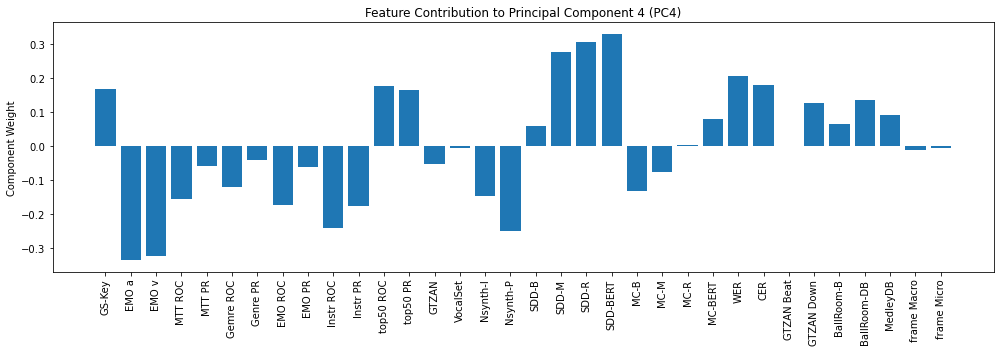}
        \label{fig:pca4}
    }

    \caption{PCA components of CMI-Bench.}
    \label{fig:mainfigure}
\end{figure}

\end{document}